# The Sun's interior metallicity constrained by neutrinos

Guillermo Gonzalez[1]*
[1]*Iowa State University, Department of Physics and Astronomy, Ames, IA 50011*

**ABSTRACT**
Observed solar neutrino fluxes are employed to constrain the interior composition of the Sun. Including the effects of neutrino flavor mixing, the results from Homestake, Sudbury, and Gallium experiments constrain the Mg, Si, and Fe abundances in the solar interior to be within a factor 0.89 to 1.34 of the surface values with 68% confidence. If the O and/or Ne abundances are increased in the interior to resolve helioseismic discrepancies with recent standard solar models, then the nominal interior Mg, Si, and Fe abundances are constrained to a range of 0.83 to 1.24 relative to the surface. Additional research is needed to determine whether the Sun's interior is metal poor relative to its surface.

## 1 INTRODUCTION

After more than 30 years of solar neutrino research, the apparent deficit of detected solar neutrinos compared to standard solar model (SSM) predictions appeared to be on the verge of resolution within the framework of MSW oscillation theory (Mikheyev & Smirnov 1985). However, as first noted by Bahcall & Pena-Garay (2004), the downward revision of the solar photospheric abundances by Asplund et al. (2005) have resulted in large discrepancies with helioseismic observations (Bahcall et al. 2005; Delahaye & Pinsonneault 2006). Specifically, the observed sound speed depth profile, convection-zone depth and convection-zone helium abundance fail to match the models. A number of possible solutions are being explored.

One class of solutions involves adjusting the physics of the solar interior, such as the radiative opacities, mixing treatment, equation of state and thermal diffusion. For example, Guzik et al. (2005) consider solar models with different diffusion treatments. They find that the adjustments required to match the helioseismic constraints are unphysically large and do not completely resolve the discrepancies. Likewise, the good agreement among the various researchers producing opacity tables for the solar interior indicate that future opacity revisions are unlikely to be very significant toward solving these problems (Badnell et al. 2005).

Another class of solutions involves adjusting the solar elemental abundances. Some elements, especially the nobel gases, are important for modeling the sun, yet are not well constrained from direct observations of the photosphere or chromosphere. For example, Bahcall et al. (2005) show that an upward adjustment of the uncertain solar Ne abundance can bring the predicted solar sound speed values into agreement with helioseismic data in the regions of the Sun's interior where they are most discrepant (from 0.4 to 0.7 $R_\odot$). They recommend an Ne abundance of $\log N = 8.29$, which is 0.4 - 0.5 dex larger than the value given by Asplund et al. (2005). This proposed fix has the advantage that it preserves the SSM paradigm by adjusting the abundance of only one element. More recently, however, Schmelz et al. (2005) and Young (2005) have determined revised solar Ne/O ratios which are more consistent with the Asplund et al. (2005) value.

A third class of solutions involves non-canonical solar models. We will focus on one early proposal – the so-called "low Z" models (Bahcall & Ulrich 1971). These are solar models with an interior metallicity reduced compared to that in the convective envelope. Although the success of the MSW effect in explaining the neutrino deficit at Earth has resulted in neglect of the low Z models, several lines of evidence have given them renewed support. For example, Fukugita & Hata (1998) calculated the most probable interior metallicity of the Sun from the measured solar neutrino fluxes. They found that the data are consistent with a homogeneous composition, but that the most probable interior metallicity is 75% that of the surface.

Yang et al. (2001) compared three solar models to helioseismic data: SSM, element diffusion and metal-enriched convection zone. Their results indicate that a model with an enriched convection zone gives similarly good agreement to helioseismic data as does a model with diffusion; both result in decreased helium abundance in the convective zone and deepening of the convection zone relative to their SSM.

A second motivation to consider low Z solar models derives from considerations of the formation and early evolution of the Solar System. Jeffery et al. (1997) estimate that the early Sun could have accreted up to 100 Earth masses of metal-rich material. The accreted material would have been thoroughly mixed throughout the convection zone, resulting in a metal-rich envelope relative to the interior. If metal-rich material is accreted after the convective envelope shrinks to near its present size (in 10 - 20 Myrs), then significant surface metallicity enhancement is possible. Ford et al. (1999) calculate that accretion of 10 Earth masses results in 0.06 dex increase in the Sun's surface metallicity, while 25 Earth masses yields an increase of 0.14 dex. The best solar spectroscopic analyses produce absolute photospheric abundances accurate to about ±0.05 dex. Thus, today it is possible to test scenarios positing accretion of at least 10 Earth masses of metal-rich material.

* E-mail: gonzog@iastate.edu

Finally, Gonzalez (2006) showed that the differences between the solar photospheric and meteoritic abundances correlate with condensation temperature. This tends to confirm accretion had taken place, since the volatiles are expected to be depleted in the accreted material. The magnitude of the differences is consistent with an envelope metal-enrichment of about 0.07 dex. The most volatile elements, such as C, N, O, Ne and Ar would not be enriched in the envelope, but Fe, which is an especially important opacity source in the core, would be enriched.

The purpose of the present study is to revisit the question of the Sun's interior metalllicity using solar neutrino rate measurements. Our method closely follows that of Fukugita & Hata (1998). In the intervening years, the Ga radiochemical experiments teams have published improved neutrino rates, and several new experimental results are available for the important $^8$B neutrinos. In addition, opacity tables have continued to receive improvement, as have other aspects of the solar models and the abive mentioned photospheric abundances. Given the continuing controversy surrounding the solar photospheric abundances, we explore below the sensitivity of the solar neutrino fluxes to changes in key element abundances.

## 2 DESCRIPTION OF METHOD

A reduction in the Sun's interior metallicity results in reduced opacity, reduced temperature and thus, reduced neutrino fluxes. The lighter abundant elements, such as C, N, O and Ne, are fully ionized in the Sun's inner core and thus have relatively less effect on the flux of $^8$B neutrinos than Fe. The individual neutrino sources have different temperature sensitivities; the solar $^8$B neutrinos originate closer to the core than the other neutrinos. To properly calculate the neutrino fluxes, then, it is necessary to know the detailed composition of the Sun.

Bahcall & Serenelli (2005) have calculated the solar neutrino flux derivatives with respect to several elemental abundances. Using these derivatives it is possible to calculate the changes in the flux from each neutrino source from changes in the abundance of an individual element. We adopt the fluxes and flux derivatives from Bahcall et al. (2005) for solar model BS05(AGS, OP), which is based on the solar abundance determinations of Asplund et al. (2005) and the Opacity Project radiative opacity calculations (Badnell et al. 2005).

In our calculations we assume that the MSW effect is an accurate description of the conversion of electron neutrinos into other flavors inside the Sun. The two parameters characterizing the MSW effect are now known with small uncertainties. We adopt the values determined by the Sudbury team (Aharmim et al. 2005): $\Delta m^2 = (8.0^{+0.6}_{-0.4}) \times 10^{-5}$ eV$^2$, $\theta = 33.9^{+2.4}_{-2.2}$ degrees. Calculation of the survival probability of the electron neutrinos also requires knowledge of their energy and of the electron density at the place of their production. The energy spectrum and radial distribution of flux for each neutrino source, along with the the radial distribution of electron density, were obtained from the late John Bahcall's web site (http://www.sns.ias.edu/ jnb/SNdata/sndata.html).

We have selected the results from three sets of neutrino experiments for the present analysis: Homestake $^{37}$Cl; GALLEX, SAGE and SNO $^{71}$Ga; and the Sudbury total $^8$B flux. The Homestake measurement is $2.56 \pm 0.23$ SNU (note, in this and other quoted neutrino fluxes, we combine the statistical and systematic errors quadratically). We adopt a weighted average of the solar neutrino fluxes from the three Ga experiments (Abdurashitov et al. 2002; Altmann et al. 2005): $69.9 \pm 4.2$ SNU. The Sudbury solar $^8$B flux measurement is $(4.94 \pm 0.42) \times 10^6$ cm$^{-2}$ s$^{-1}$; since this is a measurement of the total solar $^8$B flux, there is no need to apply the MSW correction.

The uncertainties in the theoretical solar neutrino fluxes are from Bahcall & Serenelli (2005). The adopted uncertainties for the theoretical Cl and Ga fluxes, following corrections for the MSW effect, are 0.38 and 5.0 SNU, respectively. The uncertainties in the MSW parameters have not been included in the calculations. The uncertainty of the theoretical $^8$B flux is 11.8%. These estimates do not include the uncertainty in the Fe abundance, since we treat it as a free parameter.

In all the cases presented below, we force the solar luminosity to be equal to the present value as the element abundances are changed. To maintain fixed luminosity in the models, we renormalize the neutrino fluxes using equation 2 of Bahcall (2002).

## 3 RESULTS

In the first case, we adjust only the Fe abundance. The resulting probability distribution with Fe as a parameter is shown in Figure 1 as a solid curve. In the second case, we adjusted Mg, Si and Fe together; the resulting distribution is shown in Figure 1 as a dashed curve. Since these three elements were adjusted as a group, the number of degrees of freedom was not changed in the probability calculations. The motivation for this second case is the fact that Mg, Si and Fe have similar values of condensation temperature (Lodders 2003); if Fe has been enhanced in the Sun's envelope by accretion, then Mg and Si should have been enhanced by similar amounts. Mg and Si are also important opacity sources in the Sun's interior.

The 68% bounds on Fe for the first case are 0.84 to 1.58. For the second case, the bounds on Mg, Si and Fe are 0.89 to 1.34.

The peak of the distribution for the first case occurs for an Fe abundance factor 1.15 times (+0.06 dex) greater than the surface abundance. For the second case, the peak

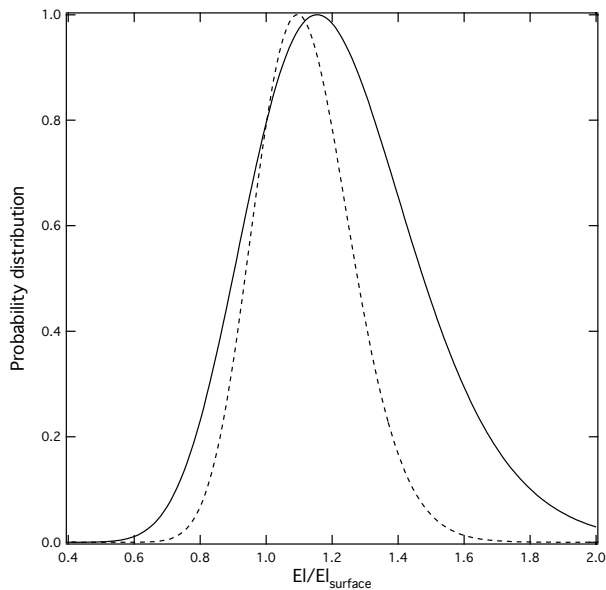

**Figure 1.** Probability distribution plotted against solar interior element abundance relative to the surface abundance. The distributions are normalized relative to the peak values. The solid curve corresponds to changing only the Fe abundance. The dashed curve corresponds to changing Mg, Si and Fe together.

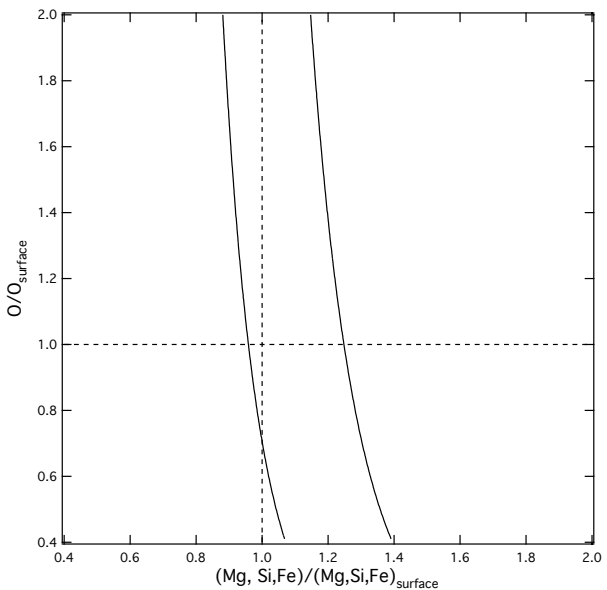

**Figure 2.** Contour diagram showing the one-sigma bounds with O and (Mg,Si,Fe) as free parameters. The fiducial surface abundances are shown as dashed lines.

## 4 DISCUSSION

We have considered four cases to test the sensitivity of the neutrino fluxes to changes in the Sun's interior composition. The measured neutrino fluxes at Earth favor a higher abundance of (Mg,Si,Fe) in the Sun's interior compared to its surface values as determined by Asplund et al. (2005), but the fluxes are still consistent with the same or even lower abundances than the surface within one-sigma.

However, recent helioseismic data are inconsistent with the abundances determined by Asplund et al. (2005). The discrepancies with the measured sound speeds can be largely resolved if the abundances of O and/or Ne are increased. As we show in Figure 2, increasing the O abundance by a few tenths of a dex will lower the (Mg,Si,Fe) abundances required to account for the observed neutrino fluxes. This solution is consistent with the abundance trend with condensation temperature in the Sun's photosphere noted by Gonzalez (2006). The observed trend implies an enhancement of the abundances of Mg, Si, and Fe by about 0.07 dex in the Sun's envelope relative to its interior.

It is important to note that low Z models predict the opposite from SSMs with respect to the Sun's interior metallicity. As a result of gravitational settling, which is included in modern SSMs, the convection zone in an evolved model should be metal poor relative to the interior by about 10% (Bahcall & Pinsonneault 1995). While gravitational settling is well understood, thermal diffusion is not. For this reason, the proper treatment of the net effect of diffusion is still rather uncertain and models with a range of diffusion treatments are still being considered to solve the discrepancies with helioseismic data.

Although the analyses presented herein are based on relatively straightforward assumptions, they could be improved upon. The evolved solar model we employed to calculate the neutrino fluxes starts with a homogenous composition and ends with a metal-rich interior due to gravitational settling.

is at 1.10 (+0.04 dex) for each element Mg, Si and Fe. Asplund et al. (2005) list the uncertainties of the photospheric abundances of Mg, Si and Fe as 0.09, 0.04 and 0.05 dex, respectively. Thus, the inferred interior Mg, Si and Fe abundances are consistent with the surface abundances to within the one-sigma errors. It is useful to note that 95% of the solutions in this second case have interior abundance greater than 0.9 of the surface abundance.

The third case is like the second except that O is increased by 0.2 dex and Ne is increased by 0.1 dex relative to Asplund et al. (2005). This results in a peak probability at 1.02 times the surface abundance; the 68% probability range is 0.83 to 1.24. The motivation for this case is the continuing controversy concerning the photospheric O and Ne abundances. As noted above, recent studies favor a Ne abundance close to the Asplund et al. (2005) value, so we are justified limiting its increase to no more than 0.1 dex. Antia & Basu (2006) use helioseismic data to show that the envelope metallicity must be higher than that determined by Asplund et al. (2005). Since O is the most abundant metal in the Sun, it is an important contributor to the overall metallicity.

The fourth and final case treats O and (Mg, Si, Fe) as two free parameters. The 68% contour lines resulting from this analysis are shown in Figure 2. This shows that the nominal surface abundances of these elements are consistent with the resulting neutrino fluxes even at the one-sigma level. The steepness of the contours confirms that a small change in the (Mg,Si,Fe) abundance can be compensated with a much larger change in the O abundance with respect to the neutrino fluxes.

An interior metallicity 0.9 times the surface value would correspond to a model with a constant abundance with depth; it is notable that our case 2 above resulted in 95% of the cases having interior abundances greater than 0.9 of the surface abundances. The calculation of the neutrino flux derivatives with respect to various abundances did not hold the surface abundances fixed; instead, the abundances were varied throughout the interior from one model to the next. A more self-consistent analysis would begin with a homogeneous model, add metals to the surface at an early stage and then evolve it to the present; the model would have to reproduce the present luminosity, radius and surface composition.

## 5  CONCLUSIONS

We find that solar neutrino flux measurements are consistent with the surface abundances determined by Asplund et al. (2005). However, helioseismic data are strongly inconsistent with SSMs built using these abundances. Recent proposals to resolve this dilemma include increasing the solar O and/or Ne abundances. This would have the effect of reducing the interior Mg, Si and Fe abundances required to account for the observed neutrino fluxes. Even without an adjustment of the surface composition, the present neutrino data are consistent with the suggestion of (Gonzalez 2006) that the Sun's convection-zone is metal enriched relative to its interior.

Although they are far from necessary to explain solar data, low Z models are still in play. There is significant motivation for reconsidering low Z models in future studies of the Sun's interior. The most immediate need is for new sets of self-consistent evolved low Z models, incorporating improved physics since the study of Yang et al. (2001). Additional progress will come from new neutrino experiments, continued comparison of helioseismic data with models having a range of element mixtures and improved measurements of O and Ne abundances in the Sun's photosphere and chromosphere.

**ACKNOWLEDGMENTS**

I thank the referee, Marc Pinsonneault, for very helpful comments.